\documentclass[12pt]{article}
\usepackage{times}
\usepackage{geometry}
\usepackage{comment}
\usepackage{floatflt}
\usepackage[utf8]{inputenc}
\usepackage{enumitem,amssymb}
\usepackage{ragged2e}
\usepackage{multicol}
\usepackage{tabularx}
\usepackage{multirow}
\usepackage{booktabs}
\usepackage{float}
\usepackage[outercaption]{sidecap} 
\usepackage{threeparttable}
\usepackage{setspace} 
\usepackage{caption} 
\newlist{thematic}{itemize}{8}
\setlist[thematic]{label=$\square$}
\usepackage{pifont}
\usepackage{graphicx}
\usepackage{tabularx}
\usepackage{sidecap}
\usepackage[utf8]{inputenc}
\usepackage[vietnamese,english]{babel}
\usepackage[dvipsnames,svgnames,x11names]{xcolor}
\usepackage{jheppub}   
\usepackage{sectsty}

\definecolor{DarkGreen}{rgb}{0.0, 0.3, 0.0}
\definecolor{purple}{rgb}{0.5, 0.0, 0.5}
\definecolor{red}{rgb}{1, 0.0, 0.0}
\definecolor{green}{rgb}{0, 1.0, 0.0}

\def\3he{$^3{\rm He}$}

\def\lsim{\mathrel{\lower2.5pt\vbox{\lineskip=0pt\baselineskip=0pt
           \hbox{$<$}\hbox{$\sim$}}}}

\def\gsim{\mathrel{\lower2.5pt\vbox{\lineskip=0pt\baselineskip=0pt
           \hbox{$>$}\hbox{$\sim$}}}}

\begin{document}
\center
\LARGE
CMB-HD: An Ultra-Deep, High-Resolution Millimeter-Wave Survey Over Half the Sky\linebreak

\vspace{0.5cm}
Response to Astro2020 Decadal Request for Information \linebreak

\Large
\vspace{0.5cm}
On behalf of the CMB-HD Collaboration \linebreak

\vspace{1.0cm}
December 9, 2019 \linebreak

\vspace{1.0cm}
\large

Neelima Sehgal$^{1}$,
Simone Aiola$^{2}$,
Yashar Akrami$^{3}$,
Kaustuv Basu$^{4}$,
Mike Boylan-Kolchin$^{5}$,
Sean Bryan$^{6}$,
Caitlin M Casey$^{7}$,
Sébastien Clesse$^{8,9}$,
Francis-Yan Cyr-Racine$^{10,11}$,
Luca Di Mascolo$^{12}$,
Simon Dicker$^{13}$,
Thomas Essinger-Hileman$^{14}$,
Simone Ferraro$^{15}$,
George M. Fuller$^{16}$,
Nicholas Galitzki$^{16}$,
Dongwon Han$^{1}$,
Mathew Hasselfield$^{17}$,
Gil Holder$^{18}$,
Bhuvnesh Jain$^{13}$,
Bradley Johnson$^{19}$,
Matthew Johnson$^{20,21}$,
Pamela Klaassen$^{22}$,
Amanda MacInnis$^{1}$,
Mathew Madhavacheril$^{21}$,
Philip Mauskopf$^{6}$,
Daan Meerburg$^{23}$,
Joel Meyers$^{24}$,
Tony Mroczkowski$^{25}$,
Suvodip Mukherjee$^{26}$,
Moritz Munchmeyer$^{21}$,
Sigurd Naess$^{2}$,
Daisuke Nagai$^{27}$,
Toshiya Namikawa$^{28}$,
Laura Newburgh$^{27}$,
\foreignlanguage{vietnamese}{Hồ Nam Nguyễn}$^{21,29}$,
Michael Niemack$^{30}$,
Benjamin D. Oppenheimer$^{31}$,
Elena Pierpaoli$^{32}$,
Emmanuel Schaan$^{15}$,
Blake Sherwin$^{28}$,
An\v{z}e Slosar$^{33}$,
David Spergel$^{2}$,
Eric Switzer$^{14}$,
Pranjal Trivedi$^{34}$,
Yu-Dai Tsai$^{35}$,
Alexander van Engelen$^{6}$,
Benjamin Wandelt$^{2,26}$,
Edward Wollack$^{14}$


\thispagestyle{empty}
\normalsize
\justify
\pagebreak
\setcounter{page}{1}
\onehalfspacing

\addcontentsline{toc}{section}{Executive Summary}
\begin{center}
\section*{Executive Summary}
\end{center}

CMB-HD is a millimeter-wave survey over half the sky, that spans frequencies in the range of 30 to 350 GHz, and is both an order of magnitude deeper and of higher resolution than currently funded surveys. CMB-HD would yield an enormous gain in understanding of both fundamental physics and astrophysics. By providing such a deep, high-resolution millimeter-wave survey (about 0.5 $\mu$K-arcmin noise and 15 arcsecond resolution at 150 GHz), CMB-HD will enable major advances. It will allow 1.)~the use of gravitational lensing of the primordial microwave background to map the distribution of matter on small scales ($k\sim10~h$Mpc$^{-1}$), which probes both dark matter particle properties and galaxy formation. It will also allow 2.)~measurements of the thermal and kinetic Sunyaev-Zel'dovich effects on small scales to map the gas density and gas pressure profiles of halos over a wide field, which probes galaxy evolution and cluster astrophysics. In addition, CMB-HD would allow us to cross critical thresholds in fundamental physics: 3.)~ruling out or detecting any new, light ($< 0.1$ eV), thermal particles, which could potentially be the dark matter, 4.)~testing a wide class of multi-field models that could explain an epoch of inflation in the early Universe, and 5.)~ruling out a purely primordial origin of galactic magnetic fields. Such a survey would also 6.)~monitor the transient sky by mapping the full observing region every few days, which opens a new window on gamma-ray bursts, novae, fast radio bursts, and variable active galactic nuclei. Moreover, CMB-HD would 7.)~provide a census of planets, dwarf planets, and asteroids in the outer Solar System, and 8.)~enable the detection of exo-Oort clouds around other solar systems, shedding light on planet formation. Finally, 9.)~CMB-HD will provide a catalog of high-redshift dusty star forming galaxies and active galactic nuclei over half the sky down to a flux limit of 0.5 mJy at 150 GHz.  The combination of CMB-HD with contemporary ground and space-based experiments will also provide countless powerful synergies. The CMB-HD survey will be made publicly available, with usability and accessibility a priority. \\

CMB-HD aims to deliver this survey in 7.5 years of observing 20,000 square degrees, using two new 30-meter-class off-axis cross-Dragone telescopes to be located at Cerro Toco in the Atacama Desert.  Each telescope will field 800,000 detectors (200,000 pixels), for a total of 1.6 million detectors. In terms of technical readiness, the CMB-HD’s detector technology and instrument design have extensive heritage from existing or already funded CMB experiments. Specifically, the technology that has already been demonstrated or is in a mature state are (1)~the dichroic, polarization-sensitive TES bolometers at 30/40, 90/150, and 220/280 GHz; (2)~the receiver design, and (3)~the baseline cross-Dragone telescope design (the latter for a 6-meter dish). These technologies are currently being fielded by the Simons Observatory (SO)~\cite{SO2019}, which will have first light in 2021, and have also been in development while planning for CMB-S4~\cite{CMBS4SB}. \\

The technology that requires further development are (1)~utilizing a laser metrology system to account for thermal effects on the 30-meter dishes, (2)~packing the specified detector elements in the receivers, which requires efficient focal plane use, (3)~scaling up existing detector and readout production and testing facilities to deliver the specified number of components within a reasonable timeframe, and (4) demonstration of large diameter ($>$ 2 meter) telescope receivers cooled to 100 mK.   Some factors mitigating these development risks are that the Green Bank Telescope (GBT) is currently testing a laser metrology system for their 100-meter dish~\cite{GBT2018}.  Recent advances in low-loss silicon~\cite{Chesmore2018} could allow a warm first lens resulting in more effective focal plane use~\cite{Niemack2016}, and an increase in observation time and/or observing efficiency can compensate for shortfalls in detector count.  Moreover, a demonstration of cooling large diameter receivers to 100 mK will be accomplished with the deployment of the SO Large-Aperature Telescope cryostat. \\

In addition, the development of MKID detector technology could substantially reduce the cost of CMB-HD.  While the baseline design of CMB-HD employs TES bolometer detectors, if MKID detectors can be robustly demonstrated to achieve comparable sensitivity to TES detectors at millimeter-wavelengths, as some indications suggest may be possible~\cite{NIKA2019}, then MKIDs could replace TES detectors resulting in a significant reduction in cost as well as technological complexity.  The BLAST-TNG, TolTEC, and CCAT-prime projects will soon advance large-format polarization-sensitive MKID arrays~\cite{Galitzki2016,Austermann2018,Parshley2018}; TolTEC and CCAT-prime will both deploy arrays at millimeter wavelengths relevant for CMB-HD in the next couple years. \\

Given that a significant amount of required technology is already in a mature state or being deployed in the field in precursor experiments in the next few years, CMB-HD could start this decade. In the first part of the decade, the technology requiring further maturity would need to be developed.  Construction of CMB-HD could then ensue in the latter part of the decade.

\clearpage
\setcounter{tocdepth}{1}
\tableofcontents

\clearpage
\begin{center}
\section{Science}

{\textbf{\textit{ 
\begin{enumerate}
\item Briefly describe the scientific objectives and the most important measurements required to fulfill these objectives. Feel free to refer to science white papers or references from the literature.
\item Of the objectives, which are the most demanding? Why?
\item Present the highest-level technical requirements (e.g. spatial and spectral resolution, sensitivity, timing accuracy) and their relation to the science objectives.
\item For each performance requirement identified, describe as quantitatively as possible the sensitivity of the science objectives to achieve the requirement.  If you fail to meet a key requirement, what would be the impact on achieving the science objectives?
\end{enumerate}
}}}
\end{center}

Two critical advances uniquely enabled by CMB-HD are mapping over half the sky:~i)~the distribution of all matter on small scales using the gravitational lensing of the cosmic microwave background (CMB), and~ii)~the distribution of gas density and gas pressure on small scales in and around dark matter halos using the thermal and kinetic Sunyaev-Zel'dovich effects (tSZ and kSZ).  The combination of high-resolution and multiple frequency bands in the range of 30 to 350 GHz allows for critical separation of foregrounds from the CMB.  That plus the depth of the survey allows one to cross critical thresholds in fundamental physics:~i.)~ruling out or detecting any new, light, thermal particles, which could potentially be the dark matter, and~ii)~testing a wide class of multi-field models that could explain an epoch of inflation in the early Universe. CMB-HD would also probe cosmic magnetic fields and open a new window on planetary studies, the transient sky, and the populations of star-forming galaxies and active galactic nuclei.  A summary of the primary science questions motivating the CMB-HD survey are given below and discussed in more detail in the accompanying science white paper and CMB-HD APC~\cite{Sehgal2019a, Sehgal2019b}. 

\vspace{-3mm} 
\begin{item}
\item $\bullet$ What is the distribution of matter on small scales? 
\item $\bullet$ What are the particle properties of dark matter?
\item $\bullet$ How did gas evolve in and around dark matter halos?
\item $\bullet$ How did galaxies form?
\item $\bullet$ Do new light particles exist that were in equilibrium with known particles in the early Universe?  
\item $\bullet$ Do axion-like particles exist?
\item $\bullet$ Do primordial gravitational waves exist from an epoch of inflation?
\item $\bullet$ If inflation happened, did it arise from multiple or a single new field?
\item $\bullet$ What is the census of bodies in the outer Solar System?
\item $\bullet$ Do Oort clouds exist around other stars?
\item $\bullet$ What is the physics behind the various bright transient phenomena in the sky?
\item $\bullet$ What are the population distributions of star-forming galaxies and active galactic nuclei?
\item $\bullet$ Was the early Universe magnetized and did that provide the seeds of galactic magnetic fields?
\end{item}

\subsection{Science Objectives}

\noindent To answer the science questions listed above, CMB-HD has the following science objectives:

\begin{enumerate}
\item {\it{\underline{Measure the small-scale matter power spectrum from weak gravitational lensing using the}}}\\{\it{\underline{CMB as a backlight; with this CMB-HD aims to distinguish between a matter power spectrum}}}\\{\it{\underline{predicted by models that can explain observational puzzles of small-scale structure, and that}}}\\{\it{\underline{predicted by vanilla cold dark matter (CDM), with a significance of at least $8\sigma$.}}} This measurement would be a clean measurement of the matter power spectrum on these scales, free of the use of baryonic tracers. It would greatly limit the allowed models of dark matter and baryonic physics, shedding light on dark-matter particle properties and galaxy evolution~\cite{Nguyen2019}. \\

\item {\it{\underline{Measure the number of light particle species that were in thermal equilibrium with the known}}}\\{\it{\underline{standard-model particles at any time in the early Universe, i.e.~$N_{\rm{eff}}$, with a $1\sigma$ uncertainty of}}}\\{\it{\underline{$\sigma({N_{\rm{eff}}}) = 0.014$.}}} This would cross the critical threshold of 0.027, which is the amount that any new particle species must change $N_{\rm{eff}}$ away from its Standard Model value of 3.04.  Such a measurement would rule out or find evidence for new light thermal particles with at least $95\%$ confidence level. This is particularly important because many dark matter models predict new light thermal particles~\cite{Baumann:2016wac,Green:2017ybv}, and recent short-baseline neutrino experiments have found puzzling results possibly suggesting new neutrino species~\cite{Gariazzo2013,Gonzalez-Garcia2019}. \\

\item {\it{\underline{Measure the primordial non-Gaussian fluctuations in the CMB, characterized by the parameter}}}\\{\it{\underline{$f_{\rm{NL}}$, with an uncertainty of $\sigma(f_{\rm{NL}}) = 0.26$, by combining the kSZ signal from CMB-HD with}}}\\{\it{\underline{an overlapping galaxy survey such as LSST.}}} Reaching a target of $\sigma(f_{\rm{NL}}) < 1$ would rule out a wide class of multi-field inflation models, shedding light on how inflation happened~\cite{Alvarez:2014vva,Smith2018,Munchmeyer2018,Deutsch2018,Contreras2019,Cayuso2018}. This cross-correlation could also resolve the physical nature of several statistical anomalies in the primary CMB~\cite{Cayuso2019} that may suggest new physics during inflation (see Ref.~\cite{Schwarz2015} for a review), and provide constraints on the state of the Universe before inflation~\cite{Zhang2015}.

\item  {\it{\underline{Remove 90\% of the CMB B-mode fluctuations from gravitational lensing over half the sky,}}}\\{\it{\underline{leaving only 10\% remaining, i.e.~achieve $A_{\rm{lens}}=0.1$.}}}  This would enable other CMB experiments with small-aperture telescopes, such as CMB-S4, to achieve their target measurement of the amplitude of primordial gravitational waves, given by the parameter $r$, with an uncertainty of $\sigma(r)<5 \times 10^{-4}$~\cite{CMBS4SB}; CMB-HD would serve as the large-aperture telescope required for B-mode ``de-lensing''.\\

\item {\it{\underline{Separately measure the density, pressure, temperature, and velocity profiles of intrahalo gas,}}}\\{\it{\underline{as a function of halo mass and redshift, by combining CMB-HD's tSZ and kSZ measurements.}}}  This would probe thermal, non-thermal, and non-equilibrium processes associated with cosmic accretion, merger dynamics, and energy feedback from stars and supermassive black holes, and their impact on the gas~\cite{Nagai2011,Nelson2014b,Lau2015,Avestruz2015,Basu2016}.\\

\item {\it{\underline{Probe the gas physics in and around halos out to $z\sim 2$ and with masses below $10^{12}$ M$_\odot$, by }}}\\{\it{\underline{stacking low-mass and high-redshift halos detected via the tSZ effect in the CMB-HD survey.}}}  The circumgalactic reservoirs of $10^{12}$ M$_\odot$ (Milky-Way-mass) halos are predicted by multiple simulations, such as EAGLE and Illustris-TNG, to be intimately linked to the appearance of, and activity within, the galaxy~\cite{schaye2015,nelson2018a,davies2019,pillepich2018}.  These and other simulations find that galactic star-formation rates, colors, and morphologies are inextricably linked not only to the mass in the circumgalactic medium, but also to the location of baryons ejected beyond $R_{200,c}$, which can be uniquely constrained by CMB-HD. Thus the science gain of such measurements is a more complete understanding of galaxy cluster astrophysics, the physics of the intergalactic and circumgalactic medium, and galaxy evolution.\\

\item {\it{\underline{Detect dwarf-size planets in our Solar System hundreds of AU from the Sun, and Earth-sized}}}\\{\it{\underline{planets thousands of AU from the Sun.}}}  In combination with  optical measurements, CMB-HD would allow large population studies of the sizes and albedos of these objects~\cite{Gerdes:2017}.\\

\item {\it{\underline{Detect exo-Oort clouds around other stars, opening a new window on planetary studies.}}\\{\it{\underline{  Advance the study of debris disks around large stellar populations. }}}}\\

\item {\it{\underline{Survey half the sky with roughly daily cadence, and make daily maps with a noise sensitivity}}\\{\underline{of 1 mJy at 150 GHz, in order to study the time variable and transient millimeter-wave sky.}}} \\The intent of the CMB-HD project is to provide to the astronomy community weekly maps of the CMB-HD survey footprint, filtered to keep only small scales, and with a reference map subtracted to make variability apparent.\\

\item {\it{\underline{Obtain a catalog of high-redshift dusty star-forming galaxies and active galactic nuclei down}}\\{\underline{to a flux limit of 0.5 mJy at 150 GHz.}}} Such a catalog would enable the study of the population distributions of these galaxies.\\ 

\item {\it{\underline{Probe the existence of primordial magnetic fields (PMFs) to find evidence for magnetogenesis}}\\{\underline{in the early Universe and reveal the seeds of observed galactic magnetic fields.}}} CMB-HD will have the high resolution and sensitivity needed to probe the magnetic vortical vector mode~\cite{Subramanian1998,Shaw2010PMF,DurrerNeronov2013,Subramanian2016} - the high-$\ell$ feature of PMFs that survives Silk damping. CMB-HD's data on magnetic signatures in non-Gaussianity~\cite{Trivedi2014,Shiraishi2014,Trivedi2012}, improved measurement of B-modes and Faraday rotation~\cite{Pogosian2018}, and measurement of the small-scale matter power spectrum~\cite{Shaw:2012} would improve magnetic field limits by an order of magnitude. CMB-HD could tighten this bound well below the 1 nG threshold, which would rule out a purely primordial origin of galactic magnetic fields~\cite{Pogosian2018}.\\

\item {\it{\underline{Constrain or discover axion-like particles by observing the resonant conversion of CMB}\\\underline{photons into axions in the magnetic fields of galaxy clusters.}}} Nearly massless pseudoscalar bosons, often generically called axions,  appear in many extensions of the standard model~\cite{PhysRevLett.38.1440,PhysRevLett.40.223,PhysRevLett.40.279,Svrcek:2006yi,Arvanitaki:2009fg,Acharya:2010zx}. A detection would have major implications both for particle physics and for cosmology, not least because axions are also a well-motivated dark matter candidate. CMB-HD has the opportunity to provide a world-leading  probe of the electromagnetic interaction between axions and photons using the resonant conversion of CMB photons and axions~\cite{Raffelt:1996wa,Mukherjee:2018oeb} in the magnetic field of galaxy clusters~\cite{Mukherjee:2019dsu}, independently of whether axions constitute the dark matter.  CMB-HD would explore the mass range of $10^{-14}$~GeV$<m_a\lesssim 2\times 10^{-12}$~GeV and improve the constraint on the axion coupling constant by over 2 orders of magnitude over current particle physics constraints to $g_{ a \gamma}<0.1\times 10^{-12}$~GeV. These ranges are unexplored to date and complementary with other cosmological searches for the imprints of axion-like particles on the cosmic density field.
\end{enumerate}

\vspace{1mm}
\subsection{Technical Requirements}\label{sec:techReq} 

\begin{table}[t]
\centering
\caption{Summary of Technical Requirements for CMB-HD Survey Over Half the Sky}
\begin{tabular}[t]{|l|ccccccc|}
\hline
Frequency (GHz)~~   & 30 & 40  & 90  & 150  & 220  & 280  & 350  \\
Resolution (arcmin)~~  & 1.25~  & 0.94~  & 0.42~  & 0.25~  & 0.17~  & 0.13~  & 0.11~  \\
White noise level ($\mu$K-arcmin)\textsuperscript{a}~~ & 6.5  & 3.4  & 0.7  & 0.8  & 2.0  & 2.7  & 100.0  \\
\hline
\end{tabular}
\begin{tablenotes}
\item \textsuperscript{a} 
Sensitivity is for temperature maps.  For polarization maps, the noise is $\sqrt{2}$ higher.
\item \textsuperscript{b} 
Range of multipoles measured is $1000 \leq \ell \leq 30,000$ for $T, E$, and $B$ maps; ~Note SO\\
will measure $30 \leq \ell \leq 1000$ to the sample variance limit for $T$ and $E$ over half the sky.
\end{tablenotes}
\label{tab:cmbhdSpecs}
\end{table}

\vspace{1mm}
The technical requirements are driven by the dark matter and new light species science targets, since those targets set the most stringent requirements.  All the other science targets benefit from and prefer these same technical requirements, since there are no science targets that pull the requirements in opposing directions.  This results in a fortunate confluence between the science targets presented above and the technical requirements that can achieve them. \\

\vspace{1mm}
\noindent {\bf{Sky Area:}} Both the dark matter and new light species science targets prefer the widest sky area achievable from the ground, given fixed observing time~\cite{Sehgal2019a,CMBS4SB}.  For the dark matter science case, the inclusion of the kSZ foregrounds is what makes wider sky areas preferable over smaller ones~\cite{Sehgal2019a}.  In addition, the non-Gaussianity inflation science target, searches for planets and dwarf planets, and probing the transient sky all benefit from the largest sky areas possible.  In practice, this is about {\it{\underline{50\% of the sky}}}, achievable from the demonstrated site of the Atacama Desert in Chile. \\ 

\begin{SCfigure}[1.4][t]
\centering
\includegraphics[width=0.55\textwidth,height=7.2cm]{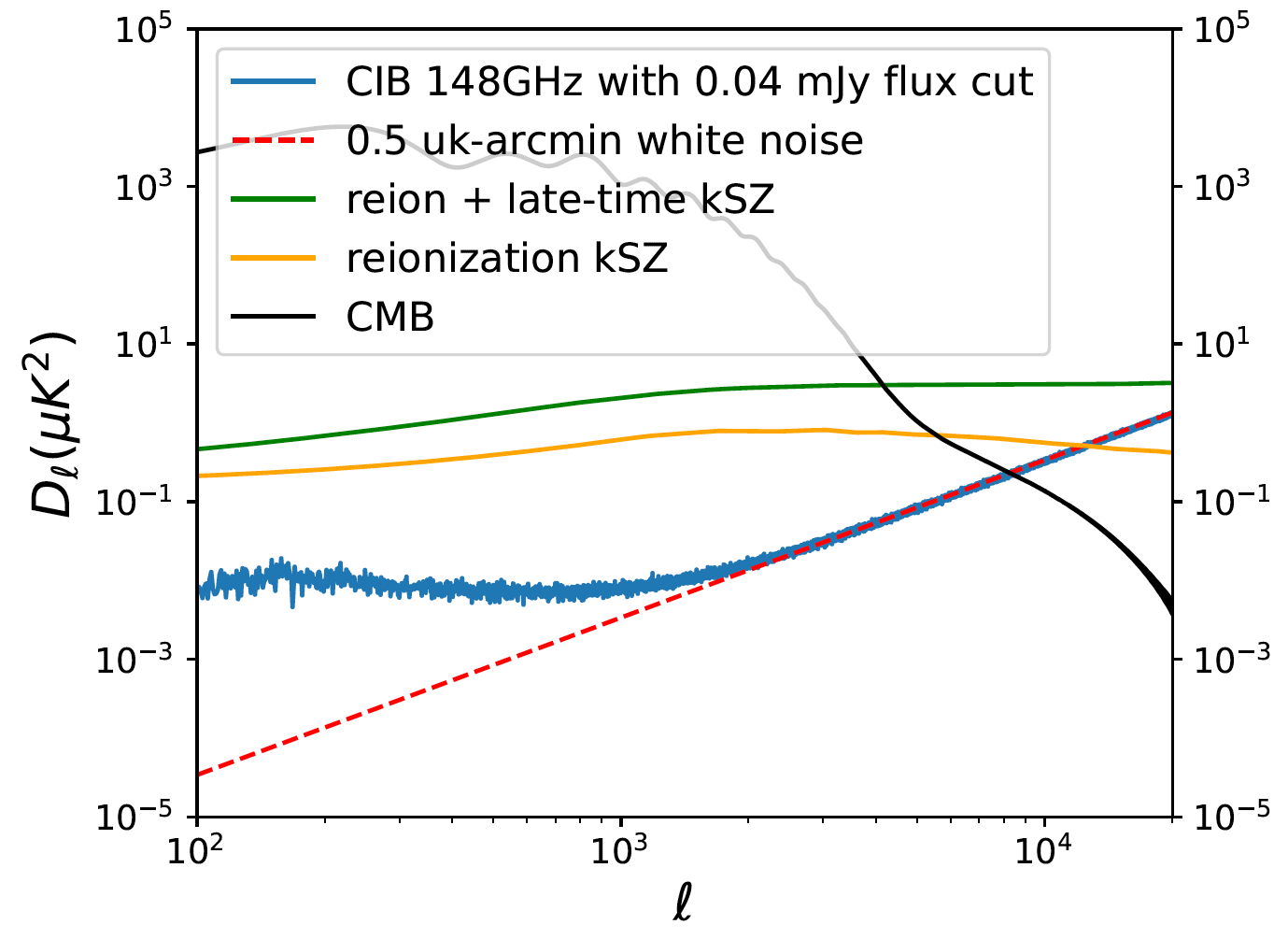}
\caption{Shown are the CMB temperature power spectrum (black solid) and relevant foregrounds at 150 GHz.  The foregrounds are the kSZ effect from the epoch of reionization (orange), reionization kSZ plus the late-time kSZ effect (green), and the CIB (after removing sources above a flux of 0.04 mJy).  The CIB flux cut, enabled by frequency channels between 100 and 350 GHz and the 30-meter dish, brings the CIB to the level of 0.5~$\mu$K-arcmin (dashed red) on small scales.}
\label{fig:foregrounds}  
\vspace{-5mm}
\end{SCfigure}

\vspace{1mm}
\noindent {\bf{Resolution:}} The resolution is set by the dark matter science target of measuring the matter power spectrum on comoving scales of $k\sim10 h$Mpc$^{-1}$ (these scales collapsed to form masses below $10^9 M_\odot$ today).  Since CMB lensing is most sensitive to structures at $z\sim 2$ (comoving distance away of 5000 Mpc), we need to measure a maximum angular scale of $\ell \sim k X \sim 35,000$. {\it{\underline{This gives a required resolution of about 15 arcseconds at 150 GHz}}} ($\ell = \pi$/radians), translating to a 30-meter telescope.  In addition, foreground cleaning will be essential, and the most dominant foreground will be extragalactic star-forming galaxies, i.e.~the Cosmic Infrared Background (CIB).  Figure~\ref{fig:foregrounds} shows that removing the sources detected at $3\sigma$ at 280 GHz with a 30-meter dish (flux cut of 0.2 mJy at 280 GHz~\cite{Lagache2018}) from the 150 GHz maps (flux cut of 0.04 mJy at 150 GHz~\cite{Sehgal2010}) lowers the CIB power at 150 GHz to below $0.5 \mu$K-arcmin.  
A resolution of 15 arcsecond is also needed to measure the profiles of the gas in halos and to separate extragalactic radio and star-forming galaxies from the gas signal.  To obtain a census of objects in the outer Solar System, we note that the parallactic motion of objects 10,000 AU away from the Sun is about 40 arcseconds in a year, also requiring this minimum resolution to detect the motion across a few resolution elements.  \\

\noindent {\bf{Sensitivity:}} The sensitivity is driven by the dark matter science target. In order to measure at the $5\sigma$-level a deviation from the CDM-only prediction of the matter power spectrum that matches claimed observations of suppressed structure, and conservatively assuming one cannot remove any of the kSZ foregrounds shown in Figure~\ref{fig:foregrounds}, requires keeping the noise subdominant to the kSZ foregrounds out to $\ell \sim 25,000$.  {\it{\underline{This requires 0.5~$\mu$K-arcmin noise in temperature (0.7~$\mu$K-arcmin in}}\\{\underline{polarization) in a combined 90/150 GHz channel.}}} Assuming removal of the late-time kSZ component with an overlapping galaxy survey like LSST, an $8\sigma$ deviation from the CDM-only prediction can be seen. 
Conveniently, this sensitivity level also allows one to cross critical thresholds, achieving $\sigma({N_{\rm{eff}}}) = 0.014$ and $\sigma(f_{\rm{NL}}) = 0.26$, as well as allows one to delens 50\% of the sky to $A_{\rm{lens}}=0.1$, a level necessary for reaching $\sigma(r)<5 \times 10^{-4}$ from a ground-based experiment. \\

\noindent {\bf{Largest Angular Scale:}} By the time CMB-HD has first light, data from the Simons Observatory (SO)~\cite{SO2019}, which will have first light in 2021, will already exist and be public.  SO will measure temperature and E-mode maps over half the sky to the sample variance limit in the multipole range of 30 to 3000 for temperature and 30 to 2000 for E-modes.  SO will also measure these scales at six different frequencies spanning 30 to 280 GHz.  Thus there is no need for CMB-HD to reimage these modes.  Therefore the largest angular scale CMB-HD needs to measure is driven by measuring B-modes, and in particular being able to remove lensing B-modes to the level of $A_{\rm{lens}}=0.1$.  {\it{\underline{This requires that the largest scale CMB-HD needs to image be about 10 arcminutes ($\ell \sim 1000)$}}}. \\

\noindent {\bf{Frequency Coverage:}} The frequency coverage is driven by needing most of the sensitivity in the frequency window that is most free from extragalactic foregrounds, namely 90 to 150 GHz.  Modern detectors can observe at two frequencies simultaneously~\cite{Henderson2016,Benson2014,OBrient2013}, so we assume we can split closely spaced frequency bands, further helping to remove frequency-dependent foregrounds.  We also require frequency coverage at 30/40 GHz to remove emission from radio galaxies and at 220/280 GHz to remove emission from dusty galaxies and to cover the null frequency of the tSZ signal (at 220 GHz).  Foreground optimization studies done for SO have found optimal ratios of noise levels given their six frequency channels, which if extrapolated to {\it{\underline{CMB-HD would require}}\\ {\underline{noise levels in temperature maps of 6.5/3.4, 0.73/0.79, and 2/4.6 $\mu$K-arcmin for the 30/40, 90/150,}}\\ {\underline{and 220/280 GHz channels respectively.}}} In addition, we require a 280/350 GHz channel in order to better exploit the Rayleigh scattering effect for improving the $N_{\rm{eff}}$ constraint~\cite{Lewis2013,Alipour:2014dza}, and in order to remove the main foreground (the CIB).  To clean the CIB to the level shown in Figure~\ref{fig:foregrounds} by removing sources above $3\sigma$ at 280 GHz
{\it{\underline{requires an additional 280 GHz channel with 3.25~$\mu$K-}}\\ {\underline{arcmin noise, for a combined noise level of 2.65~$\mu$K-arcmin at 280 GHz}}}.  This 280 GHz channel can be split into a 280/350 GHz channel at minimal cost in order to gain a 350 GHz channel with a noise level of about 100~$\mu$K-arcmin noise. In Table~\ref{tab:fluxlimits}, we give the expected single-frequency $1\sigma$ flux limits CMB-HD will achieve, in units of mJy, between 30 and 280 GHz. These flux limits are based on assuming white noise levels only, however, the confusion limit of the CIB or radio galaxies is well below these flux limits for a 30-meter dish~\cite{Lagache2018}.\\

\begin{table}[t]
\centering
\caption{CMB-HD Single-frequency Flux Limits}
\begin{tabular}[t]{|l|cccccc|}
\hline
Frequency (GHz)~~   & 30 & 40  & 90  & 150  & 220  & 280   \\
Flux limit (mJy) & 0.14 & 0.1 & 0.04 & 0.05 & 0.1 & 0.1 \\
\hline
\end{tabular}
\label{tab:fluxlimits}
\end{table}
\vspace{-3mm}
\subsection{Instrument Requirements}
\label{sec:instReq}

Given the technical requirements above needed to achieve the science targets, the following instrument specifications below are required. 

\vspace{3mm}
\noindent {\bf{Site:}} The required sky area and sensitivity make {\it{\underline{Cerro Toco in the Atacama Desert the best site for}\\{\underline{CMB-HD}}}}.  An instrument at this site can observe the required 50\% of the sky; in contrast, an instrument at the South Pole can access less than half of this sky area.  The sensitivity requirement of 0.5~$\mu$K-arcmin also requires locating CMB-HD at a high, dry site with low precipitable water vapor to minimize the total number of detectors needed.  No site within the U.S. has a suitable atmosphere.  While a higher site than Cerro Toco, such as Cerro Chajnantor in the Atacama Desert, might reduce the detector count further, that may not outweigh the increased cost of the higher site.\\

\noindent {\bf{Detectors:}} To reach the required sensitivity levels of 6.5/3.4, 0.73/0.79, and 2/4.6 $\mu$K-arcmin for the 30/40, 90/150, and 220/280 GHz channels respectively, requires scaling down the SO goal noise levels by a factor of 8~\cite{SO2019}.  SO, which is at the same site as CMB-HD, requires 30,000 detectors to achieve its sensitivity levels in 5 years of observation~\cite{SO2019}.  Since the noise level scales as sqrt$(N_{\rm{det}})$ and sqrt($t_{\rm{obs}}$), {\it{\underline{CMB-HD requires 1.3 million detectors and 7.5 years of observation to reach a}\\{\underline{factor of 8 lower noise than SO across all frequencies. An extra 280/350 GHz channel with 3.25}\\{\underline{$\mu$K-arcmin noise increases the detector count to 1.6~million.}}}}} This assumes a 20\% observing efficiency, as also assumed by SO.\\

\noindent {\bf{Telescope Dish Size:}} To achieve the required resolution of 15 arcseconds at about 100 GHz requires {\it{\underline{a telescope dish size of about 30-meters}}}.  The foreground cleaning discussed above also necessitates a 30-meter dish for the frequencies above 100 GHz, out to at least 350 GHz. \\

\subsection{Sensitivity of Science Objectives to Technical Requirements}
The impact to our science objectives of failing to meet the technical requirements are:

\begin{itemize}
\item {\bf{Sky Area:}} Surveying 50\% of the sky is most critical to constraining the existence of new light particle species (i.e.~$N_{\rm{eff}}$), and failing to meet this goal would degrade the accuracy with which we can measure this parameter. However, the sky coverage is an observational requirement, and we can adjust the survey footprint during operations to ensure that this requirement is achieved.

\item {\bf{Resolution:}} The resolution requirement is critical both for measuring the matter power spectrum on small scales and for foreground cleaning.  Achieving this resolution will thus be a key technical requirement for CMB-HD.  

\item {\bf{Sensitivity:}} The sensitivity requirement most impacts the measurement of new light particles and the matter power spectrum.  However, failure to meet the sensitivity requirement can be compensated for by increasing the observing time or potentially the observing efficency.  

\item {\bf{Largest Angular Scale:}} Failure to measure a largest angular scale corresponding to 10 arcminutes ($\ell \sim 1000)$ would only impact the ability to de-lens to a level achieving $A_{\rm{lens}}=0.1$.  This is because the next generation large angular scale CMB surveys such as the SO will be measuring temperature multipoles out to $\ell \sim 3000$ and E-modes out to $\ell \sim 2000$ to the sample variance limit over 50\% of the sky, so there is some contingency built in for T and E maps if CMB-HD does not achieve $\ell_{\rm{min}} = 1000$.  However, there is no sample variance limit for B-mode maps (in the limit of zero primordial gravitational waves), so failure to achieve $\ell_{\rm{min}} = 1000$ would degrade our ability to remove the B-mode lensing signal.  This can be compensated for to some degree by using the map of dusty star-forming galaxies measured by CMB-HD as a good tracer of the lensing potential.

\item {\bf{Frequency Coverage:}} Multiple frequencies are most critical for foreground cleaning and measuring the tSZ effect. Measuring the CMB at 30, 40, 90, 150, 220, and 280 GHz will already have been demonstrated by several CMB experiments, such as the Atacama Cosmology Telescope (ACT), BICEP/Keck, SO, Spider, and the South Pole Telescope (SPT), by the time CMB-HD begins construction. CMB-HD's 350 GHz channel is not critical for foreground cleaning, and is added as a cross check of other frequency measurements since there is minimal cost to splitting a 280 GHz channel into a dichroic 280/350 GHz channel.   
\end{itemize}

\clearpage
\begin{center}
\section{Enabling Technology}
\label{sec:enable}
\end{center}
\noindent {\textbf{\textit{Please provide information describing new enabling technologies required for activity success.  Please indicate any non-US technology required for activity success and what back up plans would be required if only US participation occurred. }}}
\begin{center}
{\textbf{\textit{ 
\begin{enumerate}
    \item For any technologies that have not been demonstrated by sub-scale or full-scale models as of this request, please describe the rationale for your technical maturity assessment, including the description of analysis or hardware development activities to date, and its associated technology maturation plan.
    \item Describe the aspect of the enabling technology that is critical to the concept’s success, and the sensitivity of mission performance if the technology is not realized or is only partially realized.
    \item Provide cost and schedule assumptions by year for all development activities, and the efforts that allow the technology to be ready when required, as well as an outline of readiness tests to confirm technical readiness level.\\
\end{enumerate}
}}}
\end{center}

\vspace{5mm}
CMB-HD is enabled by scaling up a telescope design, receiver design, and TES detector production that already has extensive heritage from existing or already funded CMB experiments.  Much of the required technology and its maturity level are documented in the CMB-S4 Technology Book~\cite{CMBS4-TechBook-2017} and CMB-S4 DSR~\cite{CMBS4-DSR-2019}, which are documents the CMB community has put together in part to give an overview of CMB technology and its readiness level.  

\begin{itemize}
\item {\bf{Telescopes:}} The telescope design for CMB-HD is largely a scaled up 30-meter version of the 6-meter cross-Dragone telescope design currently being fielded by SO and CCAT-Prime~\cite{SO2019,Parshley2018}.  There are some aspects of the 6-meter design that will need adjustment for the 30-meter version, such as a new mount design to support the additional weight.  Most importantly, CMB-HD will need to use a laser metrology system to account for thermal effects on the 30-meter dishes.  Such a laser metrology system is currently being tested on the 100-meter GBT in order to allow 90 GHz observations during the day with this telescope~\cite{GBT2018}.

\item {\bf{Detectors and Receivers:}} The baseline design of CMB-HD employs dichroic, polarization-sensitive TES bolometers at 30/40, 90/150, and 220/280 GHz.  These detectors have been used in many recent CMB experiments, both in Chile and the South Pole.  The detector technology, cooling technology, and readout technology have been successfully demonstrated in several current experiments such as ACT, BICEP/Keck, Spider, and SPT. The detector noise specifications required for CMB-HD have already been demonstrated in the field.\\

The technology that requires further development are (1)~packing the specified number of detector elements in the receivers, which requires efficient focal plane usage, (2)~demonstration of large diameter ($>$ 2 meter) telescope receivers cooled to 100 mK, and (3)~scaling up existing detector and readout production and testing facilities to deliver the specified number of components within a reasonable timeframe. For the first issue, recent advances in low-loss silicon~\cite{Chesmore2018} may make it possible to use a warm first lens sitting outside the cryostat.  This may result in a greater packing density of optics tubes within each cryostat~\cite{Niemack2016}.  Regarding cooling the large diameter receivers to 100 mK, this will be achieved with the deployment of the SO Large-Aperature Telescope cryostat in the early 2020s.  The existing facilities that successfully make and test TES detectors, such as JPL and and NIST, can scale up production.  In addition, fabrication facilities can be developed at DOE Laboratories such as Argonne and SLAC, and commercial fabrication vendors can be employed.   To mitigate the risk of potentially not achieving the specified number of detectors, we note that an increase in observation time and/or observing efficiency can compensate for this shortfall.  \\

In addition, the development of MKID detector technology could substantially reduce the cost of CMB-HD.  While the baseline design of CMB-HD employs TES bolometer detectors, if MKID detectors can be robustly demonstrated to achieve comparable sensitivity to TES detectors at millimeter-wavelengths, as some indications suggest may be possible~\cite{NIKA2019}, then MKIDs could replace TES detectors resulting in a significant reduction in cost as well as technological complexity.  The BLAST-TNG, TolTEC, and CCAT-prime projects will soon advance large-format polarization-sensitive MKID arrays~\cite{Galitzki2016,Austermann2018,Parshley2018}; TolTEC and CCAT-prime will both deploy arrays at millimeter wavelengths relevant for CMB-HD in the next couple years. \\
\end{itemize}

\clearpage
\begin{center}
\section{Telescope}
{\textbf{\textit{ 
\begin{enumerate}
    \item Provide an overview description of the characteristics and requirements of the optical telescope(s), antenna(s), or collector(s) highlighting key capabilities and any residual technology risks.
    \item Provide diagrams or drawings showing the observatory or antenna array with the instruments and other components labeled and a descriptive caption.
    \item Please provide any available review packages (e.g. Conceptual Design Review, Preliminary Design Review) that describe the scope of technical design and implementation.
    \end{enumerate}
}}}
\end{center}

\begin{figure}[bh]
\centering
\includegraphics[width=0.95\textwidth]{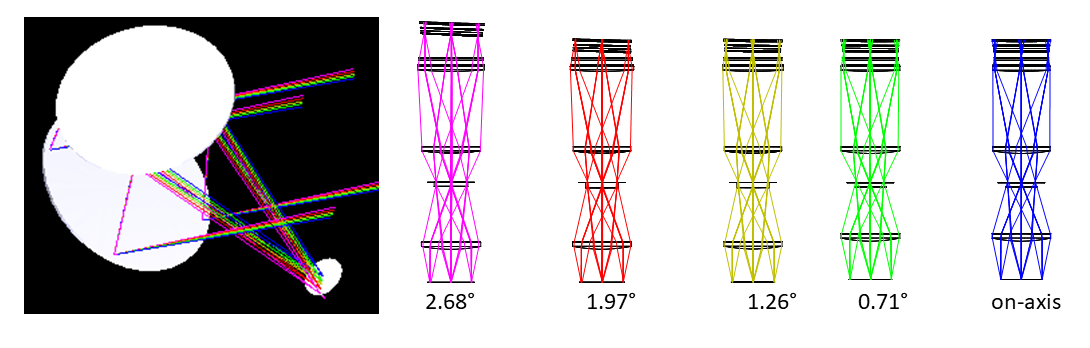}
\caption{Left: A possible design for the telescope would be a crossed Dragone design similar to that chosen for SO and CCAT-prime.  Although challenging to build, a 30-meter version of this particular design offers a diffraction limited field of view out to $r=0.8$~degrees at 150~GHz at the secondary focus, and, with cold reimaging optics such as those shown on the right, diffraction limited beams can be achieved out past a radius of 2.68 degrees at 150~GHz and even further at lower frequencies.  These particular designs are limited by the largest silicon optics currently available (45~cm) and could be grouped together in multiple cryostats in order to ensure less down time and greater flexibility.  Figure reproduced from~\cite{Sehgal2019b}.}
\label{fig:optics}  
\end{figure}

\vspace{5mm}
For systematic control, the baseline design of CMB-HD is an off-axis telescope with a primary aperature of 32 meters.  We use a crossed Dragone design because it has a larger field-of-view (fov) than Gregorian or Cassegrain telescopes.  Although the crossed Dragone design has a far larger secondary mirror (26 meters), as shown in Figure~\ref{fig:optics}, with the use of correcting cold optics extremely large fovs are possible. Simple calculations indicate that with efficient focal plane use ($>50$\%) it could be possible to fit all the detectors in one telescope.  However, given the number of receivers required, the baseline design consists of two telescopes. \\

\vspace{3mm}

\begin{figure}[]
\centering
\includegraphics[width=0.8\textwidth]{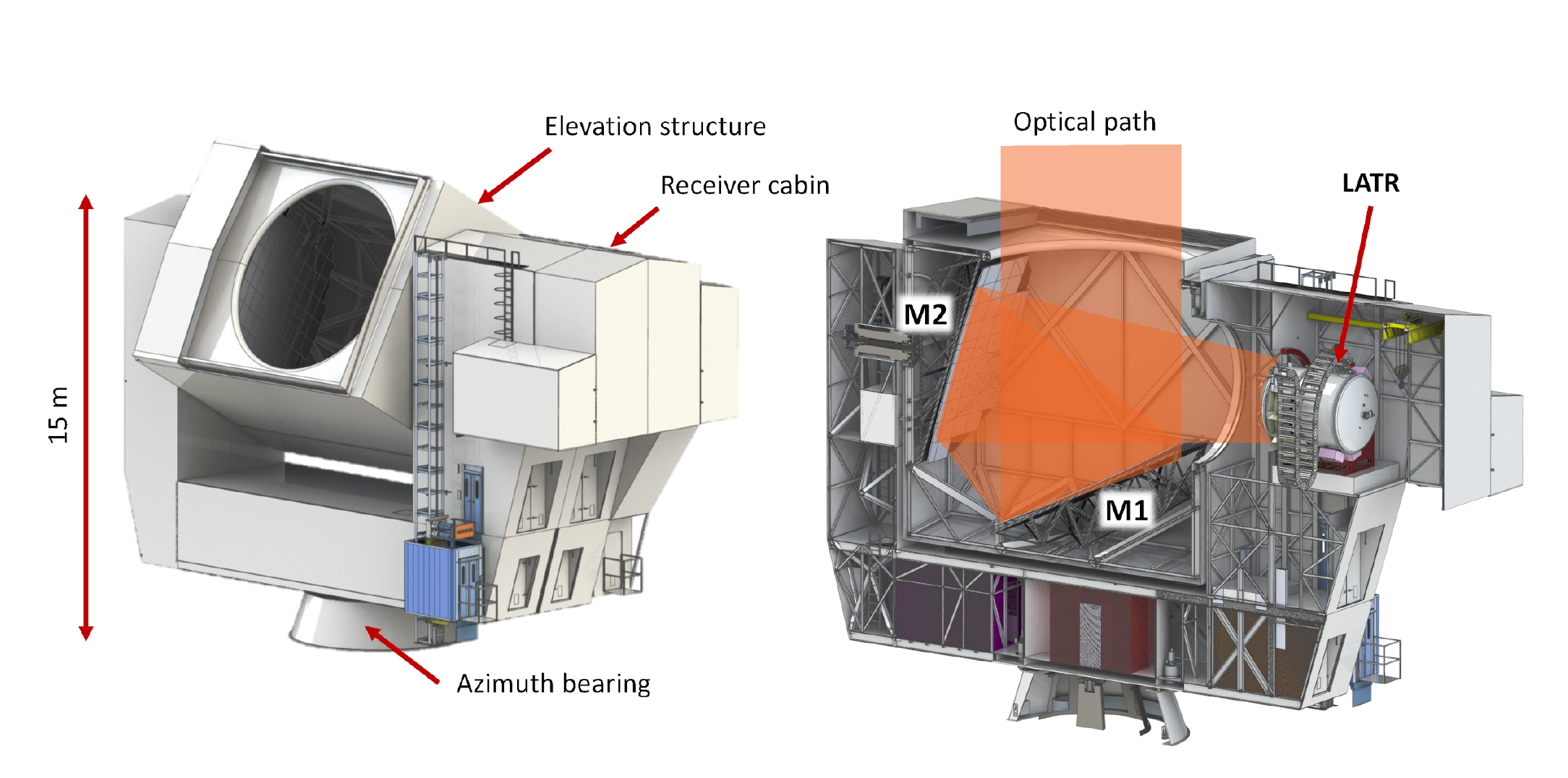}
\caption{Shown is the design for the SO large-aperture telescope (SO LAT).  The Cross-Dragone design of CMB-HD is largely a scaled up version of the SO LAT and CCAT-Prime telescope designs.  The main differences will be the mount for each CMB-HD telescope, since the CMB-HD mounts will need to support more weight, and the addition of a laser metrology system required by CMB-HD. Figure reproduced from~\cite{CMBS4DSR}.}
\label{fig:SO-LAT}  
\end{figure}

\vspace{-8mm}
\begin{center}
{\textbf{\textit{    
\begin{enumerate}
\setcounter{enumi}{3}
    \item Please describe any hardware/software with significant heritage.
    \item Please fill out the table below regarding the primary scientific equipment (e.g., Telescope or Antenna Array). Expand, contract, or modify this table as necessary and applicable.
    \item Identify and describe the three components of lowest technical maturity, and explain how and when these components will be demonstrated in hardware.
    \item What are the three greatest risks to cost, schedule, and performance?
    \item Describe any aspect of the design or implementation that may require non-US participation.
\end{enumerate}
}}}
\end{center}

The Cross-Dragone design of CMB-HD is largely a scaled up version of the SO large-aperture telescope (LAT) and CCAT-Prime designs, which are currently being manufactured (see Figure~\ref{fig:SO-LAT} for SO LAT design).  The major differences will be the mount for the CMB-HD telescopes, which will need to support more weight than the SO LAT or CCAT-Prime. Another main difference is that CMB-HD will require a laser metrology system to correct for thermal, gravitational, and wind effects on timescales of tens of seconds.  Such a laser metrology system is currently being tested on the GBT 100-meter telescope at millimeter wavelengths.  Table~\ref{tab:telescope} lists the detailed telescope characteristics of CMB-HD.  In terms of non-US participation, we note that the telescope vendor that is used by SO and CCAT-Prime is Vertex, which is a German company that makes parts of the telescopes at various locations in Europe.
\clearpage

\begin{table}[]
\caption{Telescope Characteristics Table for CMB-HD}
\label{tab:my-table}
\centering
\begin{tabular}{|l|l|}
\hline
{\bf{Optical Telescope}}	    & {\bf{Value/Summary}}	 \\
\hline
\hline
Main and Effective Aperture Size		&  32~meter actual, 28~meter illuminated            \\
\hline
System Effective Focal Length		    &    f/1.9 -- f/2.8                   \\
\hline
Total Collecting Area		            &    500--610 square meters per telescope     \\
&   \\
\hline
Field of View		                    &     $>3.2$ degree diameter at 150~GHz   \\
& (with cold reimaging optics)          \\     
\hline
Wavelength range		                & 0.86 mm to 10 mm          \\
\hline
Optical surface figure quality (RMS)	&  25~$\mu$m requirement (15~$\mu$m goal)~\textsuperscript{a}~~~~~~~~~~                   \\
\hline
Surface Coating Technique		        &   Machined aluminum               \\
\hline
Number of Mirrors or Reflecting Surfaces &    Two                \\	
\hline
Size of each Optical Element 	&    Primary: 32m diameter    \\
and its Clear Aperture & Secondary: 26~m diameter    \\	
\hline
Panel dimensions  &  Primary: 0.75 to 2.0~m  \\
& Secondary: 0.75 to 2.0~m \\
\hline
Panel surface rms (each panel)  & $<10$ microns 
\\
\hline
Mass of each Segment or Element 		&     10 kg/m$^2$ per mirror panel                \\
\hline
Total Moving Mass 	(on elevation bearing)~~~~~~~~~~	                &        700 -- 1500 tons           \\
\hline
Total Moving Mass 	(on azimuth bearing)	                &      2000 -- 7000 tons \\
\hline
Mass of each Optical Element 		    &      7000 kg mirror alone, \\ 
& 14 tons with support structure~\textsuperscript{b}~~ 
            \\
\hline
Panel actuator Precision and Range		    &     Course manual adjust +-25cm; \\
& fine (5$\mu$m precision) adjust +-1~cm~\textsuperscript{c}~~~~~~~~~~~~               \\
\hline
\end{tabular}
\begin{tablenotes}[flushleft]
\item \textsuperscript{a} Commercial laser trackers can achieve better than 10~$\mu$m measurements along a line of sight over 30 meters; a similar system is being deployed at the GBT.  We would expect such a system to correct errors in real time at a panel level taking care of thermal, gravitational, and wind effects on timescales of tens of seconds.   

\item \textsuperscript{b} 
We baseline a carbon backup structure, and can descope to the same material as the panels if thermal and cost constraints allow; a full study to minimize panel gaps while ensuring that they never drop to zero is needed and will affect the material choice. 

\item \textsuperscript{c}
The range needs to be enough to take into account all thermal and gravitational distortions with an accuracy better than our measurement of the panel location.  Lighter (more flexible but cheaper) telescope structures are possible if real-time adjustment/measurement of mirror shapes and locations are better.   
\end{tablenotes}
\label{tab:telescope}
\end{table}
\clearpage
\begin{table}[H]
\setcounter{table}{1}
\caption{Telescope Characteristics Table for CMB-HD (continued)}
\label{tab:my-table}
\centering
\begin{tabular}{|l|l|}
\hline
{\bf{Optical Telescope}}	    & {\bf{Value/Summary}}	 \\
\hline
\hline
Degrees of Freedom (mirror panels)		                & 5 actuators (corners/center) in Z               \\
& and manual x,y \\
\hline
Degrees of Freedom (mirror) & 4 (x,y,z and rotation) \\ \hline
Type of Mount used for Pointing and Allowed Range~~~~~ &  az-el, -40$<$az$<$400; 0$<$el$<$180                   \\		
\hline
Mass and Type of Material for Support Structure &   7 tons of carbon fiber (each mirror)~~~ \\
& steel telescope structure~\textsuperscript{d}~~  \\ 		
\hline
Optic Design 		    &   Cross-Dragone                  \\
\hline
Description of Adaptive Optics	    &  Laser metrology system   \\ 
\hline
\end{tabular}
\begin{tablenotes}[flushleft]
\item \textsuperscript{d}
With an active measurement system one does not need low CTE materials; steel is much more cost effective.
\end{tablenotes}
\end{table}

\clearpage
\begin{center}
\section{Instrumentation}
{\textbf{\textit{ 
\begin{enumerate}
    \item Describe the proposed science instrumentation, and briefly state the rationale for selection. Discuss the capabilities of each instrument (Inst 1, Inst 2 etc.) and how the instruments are used together. Indicate whether cryogens or other cooling are required. 
    \item Briefly describe any concept, feasibility, or definition studies already performed and please provide any available Review Packages (e.g., Conceptual Design Review, Preliminary Design Review) that describe the instrument and its design and implementation.
    \item Indicate the technical maturity level of the major elements and the specific instrument maturity of the proposed instrumentation (for each specific Inst 1, Inst2 etc.), along with the rationale for the assessment (i.e. examples of heritage, existence of breadboards, prototypes, mass/volume and power comparisons to existing units, etc. and any identifications of major long lead items). 
    \item For instrument operations, provide a brief functional description of operational modes, and calibration schemes. This can be documented in the Operations Section. Describe the level of complexity associated with analyzing the data to achieve the scientific objectives of the investigation. Describe the types of data (e.g. bits, images).
    \item Please fill out the table below regarding each instrument (if applicable). Copy as needed for all instruments. Expand this table as necessary and applicable.
\end{enumerate}
}}}
\end{center}    

\vspace{3mm}
The CMB-HD telescope cameras will hold about 400,000 pixels.  Each pixel will have two frequency bands and two polarizations for a total of 1.6~million detectors. We assume in the baseline design horn-fed TES detectors, however, MKIDS may also be a viable detector technology, which could reduce the cost significantly.  There will be four frequency band pairs:~30/40, 90/150, 220/280, 280/350 GHz.  The distribution of detectors per frequency for the first three band pairs will be similar to the ratios adopted by SO, and achieve the noise levels given above.  These ratios were calculated by calculating target map noise for each frequency from the science requirements and then taking into account how detector noise varies with frequency. \\

\begin{figure}[t]
\centering
\includegraphics[width=0.5\textwidth]{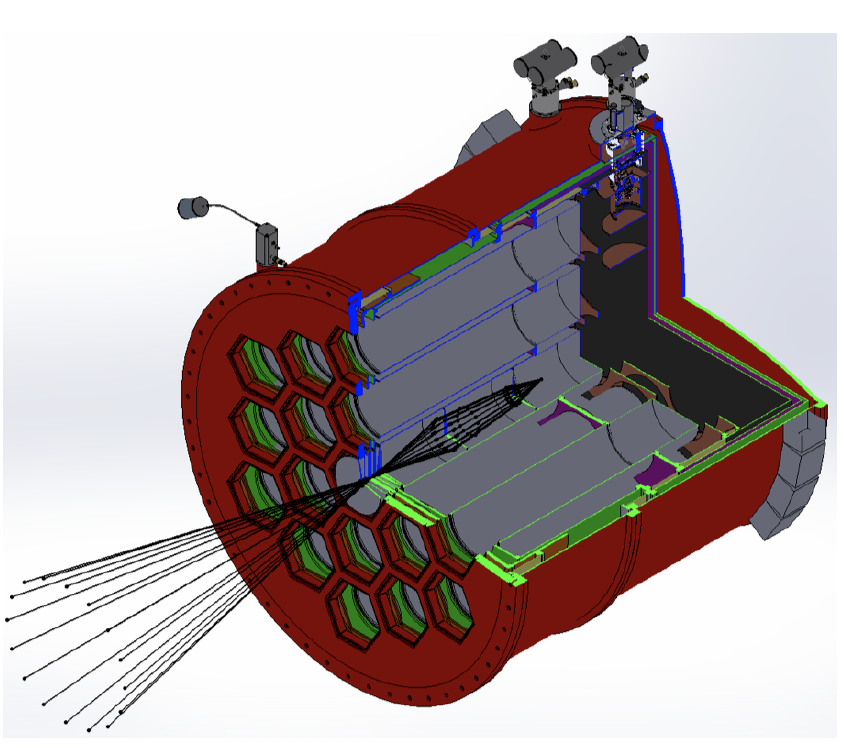}
\caption{CMB-HD will adopt a similar design for its baseline receiver as that used by SO and CCAT-Prime. Shown here is the CMB-S4 receiver design, which is an extension of the SO LATR design; the CMB-S4 receiver has a 2.6 meter diameter and 19 optics tubes, whereas the SO LATR has 13 optics tubes.  With the aid of a warm first lens, it is possible to focus light through smaller vacuum windows with very little gaps between tubes - currently tens of centimeters are needed between tubes in order to maintain the strength of the cryostat resulting in a loss of ~50\% of the telescope's focal plane. CMB-HD will aim to increase the packing density of optics tubes. Seven receivers are assumed to fit in the CMB-HD focal plane, which has a diameter of about 7.3 meters, for each telescope. Figure reproduced from~\cite{CMBS4DSR}.}
\label{fig:receiver}  
\end{figure}

For a baseline receiver we adopt a design similar to that for CCAT-prime and SO (see Figure~\ref{fig:receiver}).  Multiple sets of cold silicon lenses re-image the telescope focal plane while adding a 1K lyot stop, baffles for control of stray light, and cold blocking and bandpass filters.  For SO and CCAT-prime these are housed in a single cryostat. However, for a telescope the size CMB-HD it makes sense to group them into a number of cryostats; a single cryostat many meters across is hard to build and impractical to transport and maintain. In addition, recent advances in low-loss silicon~\cite{Chesmore2018} could allow a warm first lens, and in that case the packing density of tubes could become far greater~\cite{Niemack2016}, allowing for smaller cheaper cryostats. \\

The baseline design of CMB-HD assumes seven cryostats for each telescope, each of which has a diameter of about 2.5 meters (including flanges).  The focal plane area is about 7.3 meters in diameter, which can contain the seven cryostats.  Each cryostat holds between 13 and 26 optics tubes (depending on a final optimization of tube size versus cost and focal plane usage).  Every optics tube has a vacuum window, thermal-infrared blocking filters, and cold lenses that focus the light onto an array of detectors cooled to 0.1K.  Each optics tube will hold 500 to 4000 pixels (depending on frequency) that are sensitive to two polarizations and two frequency bands per pixel (i.e.~four detectors per pixel).  Assuming $2f*\lambda$ spacing of detectors and a conservative filling factor of the focal plane of 50\% (due to gaps between detector wafers, optics tubes, and cryostats), then with the optical throughput of our design it is possible to fit $\sim 132,000$ pixels per telescope. We will aim for a more optimistic filling factor of 75\%, yielding about 200,000 pixels per telescope.\\ 

Other than achieving the high packing density of detectors in the receivers, the assumed instrumentation is based on demonstrated technology.  Similar optics tubes have been used by ACT, BICEP/Keck, POLARBEAR, and SPT.  ACT has demonstrated successful use of horn-coupled dichroic, dual-polarization pixels.  ACT has also successfully cooled their detector arrays to 0.1K using a dilution refrigerator. Pulse-tube cryocoolers cool the optics and thermal shields to 4K.  Horn-fed TES detectors have been used for almost all recent ground-based CMB experiments, as has the time-domain multiplexing readout technology.  CMB-HD will cover a spectral range of 30 to 350 GHz.  Table~\ref{tab:inst} lists the characteristics of the receiver instrumentation.

\begin{table}[]
\caption{{} Receiver Instrumentation for CMB-HD{}}
\centering
\begin{tabular}{|l|l|}
\hline
{\bf{Item}}                                         & {\bf{Value}}      \\
\hline
\hline
Type of Instrument                           &    Polarization-sensitive bolometer cameras;            \\
& Seven cameras per telescope \\
\hline
Spectral Range                               &       Dichroic pixels at 30/40, 90/150,             \\
&  220/280, and 280/350 GHz \\
\hline
Optics Tubes &  91 to 182 per telescope,  \\
& each tube with about 40 cm clear aperture; \\
& Distributed roughly in the ratio of 1:4:2:2 for\\
& 30/40, 90/150, 220/280, and 280/350 respectively \\ 
\hline
Number of Detectors, type, and pixel count    &    1.6 million TES bolometers (400,000 pixels)              \\
\hline
Thermal or Cryogenic Requirements            &  0.1K (Pulse tubes to 4K + Dilution fridge)                \\
\hline
Size/Dimensions   (for each instrument)                           &   2.5m diameter,  2--3m long      \\
\hline
Instrument average science data volume / day &   190 TB/day uncompressed               \\
\hline
Instrument Field of View                     &   1 degree diameter (each)               \\
\hline
Development Schedule                         &   2 years design + 2 years  construction \\
& (construction schedule will be limited by \\
& detector/readout fabrication)        \\
\hline
\end{tabular}
\label{tab:inst}
\end{table}

\begin{center}
{\textbf{\textit{ 
\begin{enumerate}
\setcounter{enumi}{5}
    \item What are the three primary technical issues or risks?
\end{enumerate}
}}}
\end{center}

As we discussed in Section~\ref{sec:enable}, the primary technical risks for the receiver instrumentation are:
\begin{itemize}
\item Packing the specified detector elements into the receivers, which requires efficient focal plane use.
\item Demonstration of large diameter ($>$ 2 meter) telescope receivers cooled to 100 mK. 
\item Scaling up existing detector and readout production and testing facilities to deliver the specified number of components according to schedule.
\end{itemize}
For the first issue, CMB experiments have already demonstrated 50\% filling factors for their focal planes. The loss of focal plane is due to gaps between detector wafers, optics tubes, and cryostats.  Here we baseline a more optimistic filling factor of 75\%.  An alternative way to achieve greater packing density of detectors in a cryostat is to use a warm first lens sitting outside the cryostat~\cite{Niemack2016}.  This may be possible with recent advances in low-loss silicon~\cite{Chesmore2018}.\\

While it will be a challenge to cool large diameter receivers to 100 mK, precursor CMB experiments, such as SO, will achieve this prior to CMB-HD construction. \\ 

Regarding scaling up detector and readout facilities, the existing facilities that successfully make and test TES detectors, such as JPL and and NIST, can scale up production to some extent.  In addition, fabrication facilities can be developed at DOE Laboratories such as Argonne and SLAC, and commercial fabrication vendors can be employed.   To mitigate the risk of potentially not achieving the specified number of detectors plus readout, we note that an increase in observation time and/or observing efficiency can compensate for this loss.  

\begin{center}
{\textbf{\textit{ 
\begin{enumerate}
\setcounter{enumi}{6}
    \item Describe the heritage of the instruments and associated sub-systems. Indicate items that are to be developed, as well as any existing hardware or design heritage.
\end{enumerate}
}}}
\end{center}
As mentioned above: 

\begin{itemize}
\item The polarization-sensitive TES bolometer detectors have extensive heritage from several CMB experiments such as ACT, BICEP/Keck, and SPT.  These detectors have been sucessfully produced at Argonne National Lab, GSFC, JPL, NIST, and UC Berkeley.  Fabrication facilities at SLAC are also being developed, and commercial vendors can be employed. 

\item The time-multiplexing detector readout system has been used already by ACT, BICEP/Keck, and Spider.

\item The detailed telescope design needs to be developed, however, it will in large part be a scaled up version of existing designs.

\item The data acquisition system will be based in large part on those used for SPT-3G and the Simons Observatory.   
\end{itemize}

\begin{center}
{\textbf{\textit{ 
\begin{enumerate}
\setcounter{enumi}{7}
    \item Describe any instrumentation that may require non-US participation. 
\end{enumerate}
}}}
\end{center}
CMB-HD is open to discussions with potential international partners.  However, we currently do not assume any non-US participation.

\begin{center}
{\textbf{\textit{ 
\begin{enumerate}
\setcounter{enumi}{8}
    \item List instruments to be delivered as part of construction versus ongoing development when in operations.
\end{enumerate}
}}}
\end{center}
The expectation is that all CMB-HD instrumentation will be delivered as part of construction, and not during operations.

\begin{center}
{\textbf{\textit{ 
\begin{enumerate}
\setcounter{enumi}{9}
    \item Discuss anticipated data rates and volumes, as well as plans for processing and archiving data.
\end{enumerate}
}}}
\end{center}
The anticipated data rates and volumes will be significantly larger than any precursor CMB experiment.  We anticipate 17 Gbps or 2.24 GB/sec as the instant, uncompressed data rate. Compression will reduce this by a factor of a few.  This yields 190 TB/day or 67 PB/year.  Over the 7.5 year survey, we anticipate 505 PB of data.  While storing so much data is costly today, we can expect that in a decade storing this data volume will be much more feasible.  We would plan to archive the data at multiple high-performance supercomputing sites, such as the DOE's NERSC and ALF centers, as well as NSF's XSEDE computing center.

\clearpage
\begin{center}
\section{Facilities}
{\textbf{\textit{
\begin{enumerate}
\item Describe the site and its location, including size, altitude, access, number of buildings, size of building(s) footprint and volume, existing infrastructure, power, internet, environmental considerations and logistics (proximity to major airport, housing and support for construction crews and facility staff etc.).
\item Identify which facilities will be new and which facilities may be pre-existing. Describe any existing facilities and their estimated remaining useful life. Describe any upgrades to existing facilities that will be undertaken. Describe any anticipated shared use of site facilities between the concept being proposed and existing telescopes.
\item For antenna arrays, provide specific infrastructure required such as concrete pad size, communications buildings, etc. for each element in the array. For telescope mirrors, describe infrastructure needed for mirror maintenance, e.g. coating facilities.
\item Describe atmospheric and radio frequency interference (RFI) characteristics of the site insofar as they would affect observations with the concept being proposed.
\end{enumerate}
}}}
\end{center}

\vspace{5mm}
Cerro Toco in the Atacama Desert is the best site to achieve the science of CMB-HD.  An instrument at this site can survey the required 50\% of the sky.  The sensitivity requirement of 0.5~$\mu$K-arcmin also requires locating CMB-HD at a high, dry site with low precipitable water vapor to minimize the total number of detectors needed.  No site within the U.S. has a suitable atmosphere.  While a higher site than Cerro Toco, such as Cerro Chajnantor in the Atacama Desert, might reduce the detector count further, that likely does not outweigh the increased cost of the higher site.  \\

\begin{figure*}[th]
  \centering
  \includegraphics[width=0.80\textwidth]{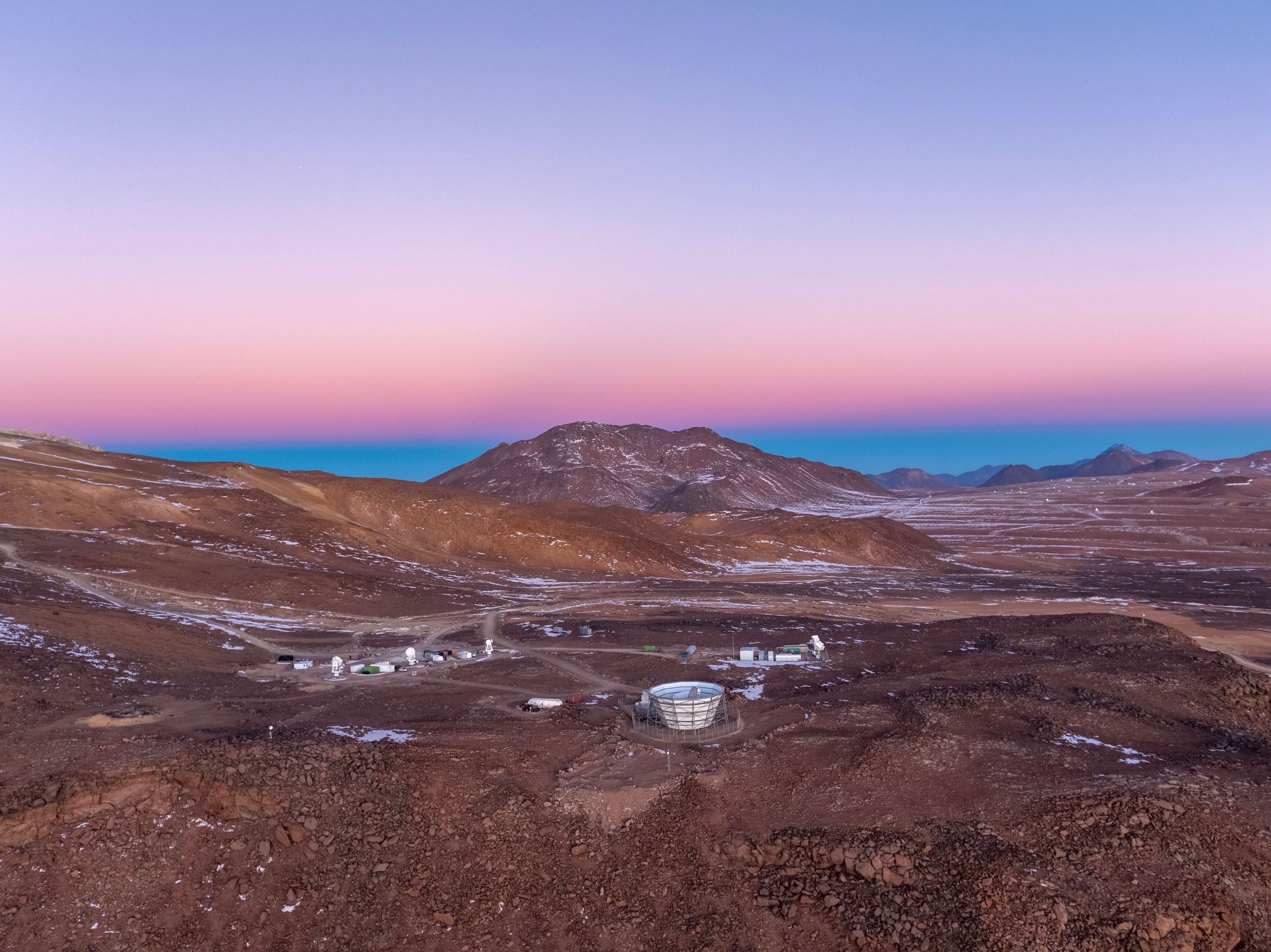}
  \includegraphics[width=0.80\textwidth]{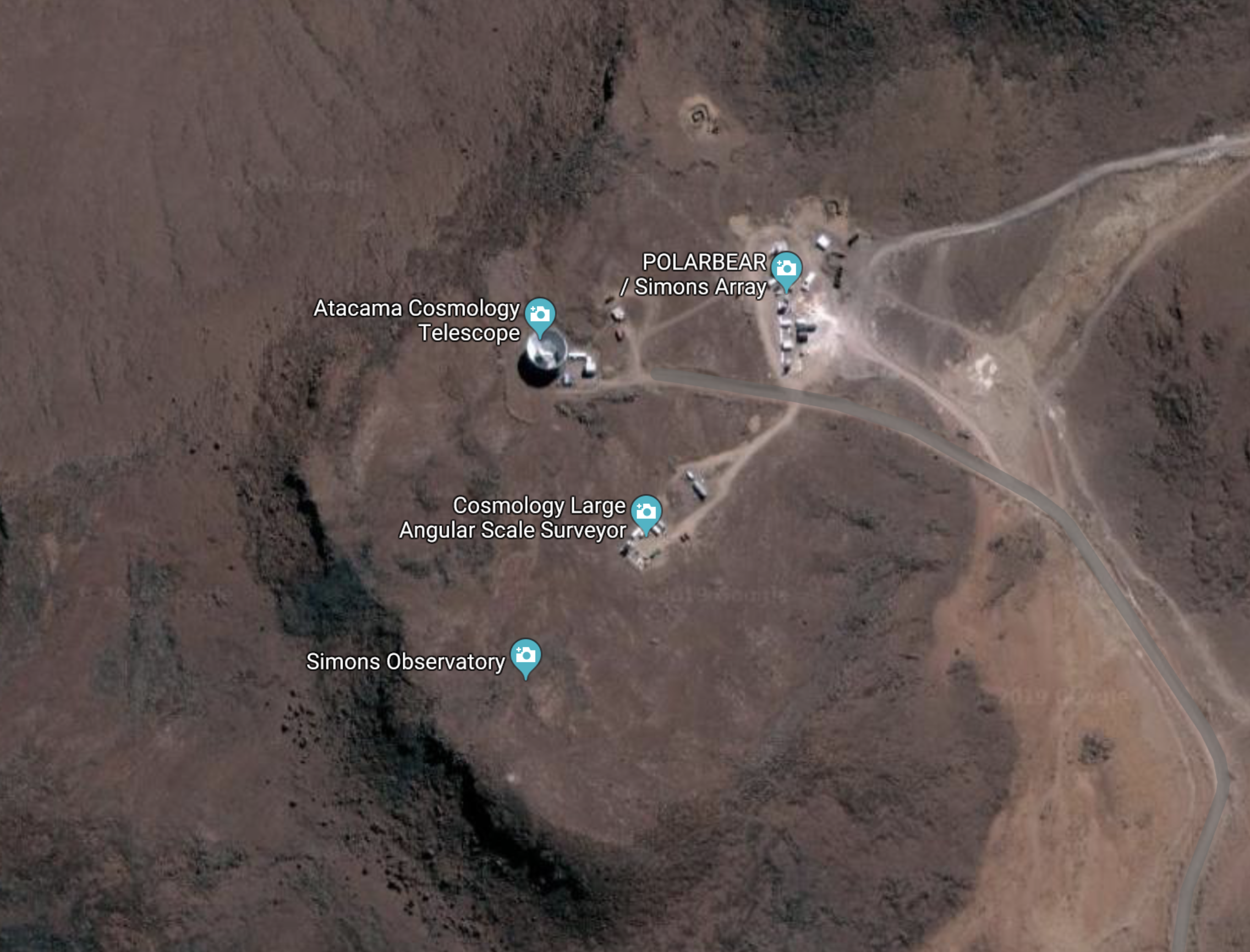}
  \caption{Cerro Toco in the Atacama Desert. {\it{Top:}} Photo courtesy of Debra Kellner. {\it{Bottom:}} Image from Google Earth.  Telescopes pictured are the Atacama Cosmology Telescope (ACT), POLARBEAR/Simons Array, and the Cosmology Large Angular Scale Surveyor (CLASS). Also shown is the site of the funded Simons Observatory (SO).}
  \label{fig:site}
\end{figure*}

Figure~\ref{fig:site} shows the existing CMB telescopes already on this plateau, namely ACT, POLARBEAR/Simons Array, and the Cosmology Large Angular Scale Surveyor (CLASS). Also shown is the location of the funded SO, which is expected to have first light in 2021.  In addition to the footprints of these existing and planned telescope facilities, there is about 300,000 square feet of available level ground at the Cerro Toco site that is appropriate for supporting telescopes and accompanying buildings.  SO will build two new buildings at this site, one that includes a high-bay lab (11m x 8m x 8.5m high) that can support an 8-ton crane, as well as one that includes a control room and office. These buildings are designed to satisfy engineering code IBC-2009, with maximum allowed winds of 90 mph. CMB-HD would build at least two new buildings, roughly similar in scope to the SO buildings.  \\

\clearpage
The site itself is 5200 meters above sea level. It sits on public Atacama Astronomical Park (AAP) land.  From the nearby town of San Pedro de Atacama, small trucks can reach the site in about 50 minutes via public roads. Large trucks bringing deliveries take about 90 minutes travelling to the site on longer ALMA-controlled roads. In general, people working at the site will travel there daily from the town of San Pedro.  There are a number of facilities in San Pedro capable of providing food and housing.\\

In order to use the AAP land, one must satisfy a biological study and an archaeological study.  If these studies find no biological or archaeological items that needs protection, then construction can commence as long as an archaeologist is present during excavation to ensure nothing of historical interest is discovered. None of the existing CMB experiments at the site have ever encountered anything preventing construction.\\

In general, power at the site is a solved issue with a baseline of either diesel or gas turbine generators. Significant savings can be had with supplementary solar power during daylight hours which would pay for itself in a few years of operations while also lowering our carbon footprint.  We expect that the total power requirements of CMB-HD can be met with a combination of solar array, battery power, and diesel generators.  CMB-S4 has estimated a cost of \$10 million dollars to install a hybrid photovoltaic/battery/diesel power plant on site~\cite{CMBS4RFI}. \\

Internet access to and from the site is provided by the National Research and Education Network of Chile (REUNA).  REUNA has a local access point about 10 kilometers away.  SO is investing in infrastructure to connect SO to this REUNA access point.  SO has achieved a data rate of about 1 Gbps from this local access point to the DOE National Energy Research Scientific Computing Center (NERSC).  However, the fundamental limit of data transfer is likely larger (between 10 and 100 Gbps), with higher bandwidth costing more~\cite{CMBS4RFI}.  \\

The nearest major airport to San Pedro is the El Loa Calama Airport (CJC), which is about 100 kilometers and 1 hour and 15 minutes away via a major highway.  There are multiple flights a day to CJC from either Santiago in Chile or Lima in Peru.  Shuttles run from CJC to San Pedro after each flight arrives.  Personnel travelling to San Pedro from the United States typically take about 20 hours for the commute.\\

Cargo sent to the site via air transportation usually go through Santiago airport and then are brought to the site via truck.  Cargo sent by ship enter Chile via ports either near Valparaiso or Antofocasta; they then are transported to the site via truck.  \\

Currently fuel is delivered to the site via 5000 L trucks to support exisiting CMB experiments located there.  For larger fuel deliveries the access road would need some improvement.  \\

CMB-HD may share the existing REUNA local access point for internet connection, as well as the planned SO fiber connection, which will have enough bandwidth for all current and planned CMB ground-based experiments based in Cerro Toco.  CMB-HD may also share the existing road access to the site, and share the facilities within the town of San Pedro for providing room and board to personnel.  \\

The foundation for each of the two telescopes has not been fully designed yet.  The most stringent criterion the design must meet is seismic survival.  \\

Regarding the atmospheric characteristics of the site, Cerro Toco is one of the driest places on Earth, which is a necessity to achieve the science of CMB-HD.  This site has high atmospheric transmittance and low atmospheric emission across the relevant millimeter-wave frequencies; this has been studied extensively as discussed in~\cite{Bustos2014, Lay2000}.  The Cerro Toco site has been used for over a decade by ACT and POLARBEAR/Simons Array.  As a result, how the atmosphere impacts the noise properties of existing instruments has been well quantified, and this information was propagated into the detector requirements for CMB-HD discussed in Section~\ref{sec:instReq}. \\

External radio frequency interference (RFI) has so far not been an issue for any existing CMB experiments at this site.  These existing CMB experiments currently monitor the RFI characteristics of the site at frequencies spanning 30 to 290 GHz, basically the full CMB-HD spectral range. In order to use the AAP land, it will be necessary to study any possible new RFI signal from a new CMB experiment planned for the site, and to assess the impact of that on any existing experiments. Moreover, ALMA continuously monitors the RFI characteristics of the area.

\clearpage
\begin{center}
\section{Operations and Observation Strategy}
{\textbf{\textit{
\begin{enumerate}
\item Please provide operations plans or documents (Concept of Operations). 
\end{enumerate}
}}}
\end{center}

   The annual operations for CMB-HD include project management, staff at the site, maintenance of the instrument, utilities, and transmission of data.  Also included are the core science analyses and delivery of core data products.    

\begin{center}
{\textbf{\textit{
\begin{enumerate}
\setcounter{enumi}{1}
    \item Provide a description of the facility operations. For example: number of staff required, position types, and 24-hour operation requirements.
    \end{enumerate}
}}}
\end{center}

The telescopes and cameras will be collecting data continuously, however, almost all functions will be automated so that local operators are not required during standard operations.  ACT, SO, CCAT-prime, and many other projects have extensive experience operating remotely at similar or higher sites in Chile.  CMB-HD will follow a similar model with a permanent presence of a small team (3 to 4) in Chile backed up by remote observers from around the world.  The remote observers will consist of shifts of two people, who will monitor in real-time the quality of the data, the weather, and if there are any issues that need to be addressed immediately.  In the latter case, the remote team will send out alerts to local staff near or at the site so that the issue can be dealt with.  A much larger team would be present in Chile for initial construction and testing.

\begin{center}
{\textbf{\textit{
\begin{enumerate}
\setcounter{enumi}{2}
    \item Provide a description of the science operations. For example: pre-observational planning, data processing/reduction required, coordination with other facilities.
    \end{enumerate}
}}}
\end{center}
The science operations will proceed as follows. 
\begin{itemize}
\item The raw data will be transferred to supercomputing centers, such as NERSC, in real-time over optical fiber. 

\item A data management team will analyze the raw time-ordered data and turn them into maps of the sky at each observing frequency that are calibrated and well characterized.  This will also require the use of one or more supercomputing centers.  In addition, this team will provide ancillary data such as well-characterized beams and frequency bandpasses.  

\item A team of specialized CMB data analysts analyze the maps to provide the vast array of scientific data products enabled by CMB-HD.  The CMB-HD collaboration consists of many such specialized CMB data analysts.  This team also informs the data management team of residual systematic effects in order to improve the mapmaking.  Standard tools to take power spectra of the maps or make lensing potential maps have already been developed by the CMB community.  In addition, SO is developing an array of additional science analysis software tools that it intends to make public to the wider CMB community.   We note that the full cost of the nearly infinite possible science analyses are not included in the baseline cost of CMB-HD, as is the custom with CMB experiments.  It has been common practice for CMB data analysts that are part of a CMB collaboration to obtain external funding to support any particular, non-core high-level analyses that they would like to perform on the data. 

\item The core project deliverables, after a modest proprietary period, will be made public to the broader scientific community via NASA's Legacy Archive for Microwave Background Data Analysis (LAMBDA).  In addition, core deliverables can be placed in the CMB community repository at NERSC.  These deliverables include (a)~well-characterized maps at each frequency from which the broader science community can derive a range of science results, (b)~power spectra, covariance matrices, and likelihood functions, and (c)~Monte Carlo simulation sets of the data.  The project also intends to provide the astronomy community with (d)~weekly maps of the CMB-HD survey footprint, filtered to keep only small scales, and with a reference map subtracted to make variability apparent.
\end{itemize}

\noindent Pre-observational planning includes developing and having in place the necessary data reduction software pipelines to go from time-ordered data to maps.  In addition, software pipelines for foreground cleaning, obtaining power spectra, and obtaining lensing potential maps should be in place prior to the start of observations.  Moreover, Monte Carlo simulations that model both the sky and the experiment should be in place close to the start of operations, however, they will be refined as the specific noise characteristics of the instrument become better known.  These simulations will be vital for verifying analysis pipelines as well as quantifying uncertainties.  

\begin{center}
{\textbf{\textit{
\begin{enumerate}
\setcounter{enumi}{3}    
    \item Discuss observatory efficiency, e.g. the impact of maintenance and engineering and calibration time on faction of science time availability.
    \end{enumerate}
}}}
\end{center}

ACT's effective observational efficiency has been about 20\%, and we assume this conservative efficiency in the CMB-HD sensitivity forecasts and detector counts assumed in Sections~\ref{sec:techReq} and~\ref{sec:instReq}. This 20\% efficiency assumes only nighttime observations during the good observing season are used, which is a small fraction of the total possible observing time.  It also includes time lost due to power outages, calibration, and various tests.  With some investment this number could be increased, for example, by using the laser metrology system to open up usability of day-time observations.
 
\begin{center}
{\textbf{\textit{
\begin{enumerate}
\setcounter{enumi}{4}    
    \item Describe scope of engineering activities and time (day or night) needed to maintain calibration and health of telescope/array and instruments.
    \end{enumerate}
}}}
\end{center}    

Routine maintenance of the facilities would occur year round during the day.  These activities include replacing and lubricating mechanical parts, inspecting and replacing electrical cabling, insulation, and temperature regulation hardware, maintaining the cryostats, and upgrading computer components and software.  Any major maintenance would take place during the summer when the weather is not good for observations.

\begin{center}
{\textbf{\textit{
\begin{enumerate}
\setcounter{enumi}{5} 
    \item Summarize key software development and any science development required.
     \end{enumerate}
}}}
\end{center}

The data management team and core science analysis team will need to have software pipelines in place to go from time-ordered data to maps to core project deliverables.  This will require pre-observation development and verification of software on simulated observations.  There will be a substantial body of public CMB software to build from prior to construction of CMB-HD. \\

Since CMB-HD will open a new frontier of observations never achieved before, in terms of the combination of sensitivity, resolution, and sky coverage, a host of new science analyses and their corresponding software pipelines will also be developed. This is happening already, as CMB theorists/data analysts are developing new science analyses that can be achieved with CMB-HD data.  They are forecasting the new measurements CMB-HD can deliver, and showing precisely how CMB-HD can push back the boundaries of our knowledge~\cite{Lague2019,Mukherjee2019,Zorrilla2019,Chung2019,Coulton2019}. Putting these high-level analyses into practice using real data presents additional challenges, however, the Simons Observatory is providing a good model for engaging CMB theorists and data analysts in software pipeline development and code verification on a common set of simulations.  CMB-HD will follow a similar model.

\clearpage
\begin{center}
{\textbf{\textit{
\begin{enumerate}
\setcounter{enumi}{6} 
    \item Summarize any archiving requirements. 
     \end{enumerate}
}}}
\end{center}

The CMB-HD archiving requirements will ensure that there is secure, long-term storage of all raw data products from the cameras.  This can be achieved by the substantial archiving facilities at NERSC as well as at the DOE Laboratories such as ALCF at Argonne National Lab.  In addition, all legacy data products and core project deliverables will be archived using the well-established facilities provided by LAMBDA.  
    
\begin{center}
{\textbf{\textit{
\begin{enumerate}
\setcounter{enumi}{7}     
    \item Describe any high-level safety policies that will be required for items of particular safety concern.
     \end{enumerate}
}}}
\end{center}

New regulations in Chile restrict work at high altitudes, and CMB-HD will follow these new regulations, adopting similar safety policies as other observatories in the area.  These policies include the requirement that multiple people be on site at the same time.  The policies also regulate the use and availability of oxygen.  In addition, the high-elevation safety regulations require a safety officer to be present on-site during major construction activities.   Transport to and from the site has also been a cause of accidents in the past, and we intend to pay careful attention to this safety issue.

\clearpage
\begin{center}
\section{Programmatic Issues and Schedule}
{\textbf{\textit{
\begin{enumerate}
    \item Please provide a programmatic overview that describes the structure of the overall organization including any international partners or university partners etc. and any money or hardware they are providing. Clearly indicate schedule and costs to date highlighting what has already been delivered and/or clearly capturing progress to date, if applicable.
\end{enumerate}
}}}
\end{center}

The CMB-HD concept was originated in the summer of 2018.  In December of 2018, we held a multi-day CMB-community workshop at the Computational Center for Astrophysics in the Flatiron Institute to flesh out the science case of, and the instrumental path towards, an ultra-deep, high-resolution CMB survey over a large fraction of the sky. The product of this workshop was a science white paper submitted to the Astro2020 Decadal in March of 2019~\cite{Sehgal2019a} that outlined the compelling science case that could uniquely be achieved by a CMB-HD survey.  Following this, in July 2019, we submitted an APC white paper to the Astro2020 Decadal~\cite{Sehgal2019b}, which described the CMB-HD project in more detail, including the flowdown from science goals to technical requirements, and the instrumentation necessary.  To date CMB-HD's newly formed collaboration consists of about 50 collaboration members.  Many in our collaboration will be engaged in another CMB-community-wide month-long workshop that will take place in Aspen during the summer of 2020, which will be focused on exploring the new discovery potential in the era of high-resolution, low-noise CMB experiments.  We note that the CMB-HD collaboration is open to the entire U.S. science community.  International partners are also welcome.  

\vspace{-6mm}
\begin{center}
{\textbf{\textit{
\begin{enumerate}
\setcounter{enumi}{1} 
    \item Please describe any funding availability challenges and the total impact it had, or may have, to cost and schedule.  Has all required funding been committed?  If not, please explain.  Highlight in-kind contributions in the past and planned from partners.  Clearly describe what the NSF/DOE/Fed Gov funding request is and what it buys.
    \item Please describe any unexpected challenges, key risks, and/or current outstanding risks that require(d) significant mitigation which affect/affected cost \& schedule.
    \item Describe the current top 3 risks to the development of the facility, and proposed mitigation strategies. Please provide a top-level schedule.
\end{enumerate}
}}}
\end{center}

To date, the CMB-HD project has not requested any funding from any organization.  For a project of such a large scale it is important to have the support of both the astronomical and particle physics communities, as well as both DOE and NSF funding agencies.  While the CMB-HD project is not ``shovel-ready'', with a few years of design and technology development, it could be ready for construction in the latter half of the decade.  It is our understanding that the DOE HEP Cosmic Frontier program is interested in a new flagship project, which would phase in as the LSST and DESI projects move into an operations phase.  The DOE has also made clear that they are open to this new project being focused on CMB measurements, as recommended by the 2014 Particle Physics Project Prioritization Panel (P5).  The timing is thus potentially ideal to support CMB-HD as a joint partnership between NSF-AST, NSF-PHY, and DOE-HEP.  A strong endorsement and recommendation by the Astro2020 Decadal would allow the CMB-HD project management to approach the DOE and NSF funding agencies, having support from the greater Astronomical community.  \\

In terms of what federal support for CMB-HD would buy, the unprecedented sensitivity and resolution of CMB-HD would open up vast uncharted territory in the millimeter-wave sky, and is a clear leap beyond currently funded ground-based CMB experiments.  It would allow us to cross critical measurement thresholds and definitively answer pressing questions in astrophysics and fundamental physics of the Universe.  The CMB-HD project also recognizes the productivity benefit of an engaged collaboration, and is committed to fostering a culture that enables this.  The CMB-HD survey will be made publicly available, and the project will prioritize usability and accessibility of the data by the broader scientific community.\\

The CMB-HD collaboration also recognizes that there are synergistic opportunities between CMB-HD and CMB-S4~\cite{CMBS4SB}.  The combination of the 18 small-aperture telescopes located at the South Pole, as proposed by the CMB-S4 project, and the two 30-meter telescopes located in the Atacama Desert, as proposed by the CMB-HD project, would be a win-win scenario for both collaborations.  The primordial gravitational wave science goals of CMB-S4 would be achieved with the 18 small-aperature telescopes measuring large-scale CMB B-modes, plus the CMB-HD large-aperture telescopes, which would be needed to separate the primordial B-mode signal from the B-mode signal induced by the gravitational lensing of CMB E-modes. All other CMB-S4 science goals would be achieved by CMB-HD, as would, in addition, the compelling science goals uniquely enabled by CMB-HD.  We also recognize that there exists common ground with the AtLAST project proposal~\cite{Bertoldi2018}, and potential international partnership.  We further note that CMB-HD can achieve many of the science goals of the original CCAT 25-meter proposal~\cite{CCAT}, which was highly ranked in the Astro2010 Decadal but not pursued.  Although CMB-HD does not extend as high as CCAT into the submillimter, with 350 GHz being the highest channel, CMB-HD's much higher sensitivity as compared to CCAT uniquely enables the science described above.

\clearpage
\begin{center}
\section{Cost}
{\textbf{\textit{
\begin{enumerate}
    \item Provide a high-level facility unique Work Breakdown Structure (WBS) with definitions. 
    \item Fill out the cost estimate tables below using the broad bins provided.  Note the tables are divided into US Only and with International Partners.  
    \item Provide a basis of estimate for project WBS elements at the highest levels (i.e., systems and major subsystems, as indicated in the cost tables below), including internal overhead rates that are applied to costs for labor.  Describe approach or methodology for highest cost element estimates such as analogies, models, expert judgement, construction bids, etc.  Describe cost assumptions made for each element such as any production learning curves, labor and major procurements.  Identify major subcontracted procurements and associated vendors if available.  The “Prior” column is meant to be actual costs incurred. 
\end{enumerate}
}}}
\end{center}

\vspace{5mm}
Table~\ref{tab:wbs} provides the CMB-HD WBS at Level 2 with definitions.  The CMB-HD instrument and site cost is shown in Table~\ref{tab:cost} with 2019 estimates. Much of the cost is in fabricating the detectors and their readout electronics.  For detector cost comparison, CMB-HD will have a factor of 3.2 more detectors than the proposed CMB-S4 experiment (1600k for \$320 million versus 500k for \$100 million).  However, there are two avenues that could potentially reduce this cost.  One is that over the next several years, as upcoming experiments such as SO require tens of thousands of detectors, the mass production of detectors may drop the cost.  The other is that MKIDS, a detector technology alternate to TES devices, are currently being tested on-sky at millimeter wavelengths~\cite{Austermann2018}. If MKID detector technology matches in practice to its promise, then the detector cost may drop by a factor of a few.  \\

Two 30-meter-scale off-axis, crossed Dragone telescopes are costed at \$400 million. The telescope receivers are estimated at \$100 million (roughly a factor of 3 higher than estimated for CMB-S4).  Project management is estimated at \$10 million per year over ten years, and site infrastructure is estimated at \$50 million total. Data management and data acquisition are estimated at 12 FTEs/year and 4 FTEs/year, respectively, assuming \$200,000 per FTE (including 40\% for benefits and 60\% for overheads). \\

\begin{table}[H]
\caption{{} CMB-HD WBS at Level 2 with Definitions {}}
\label{tab:my-table}
\centering
\begin{tabular}{|l|l|}
\hline
{\bf{Work Breakdown Structure}}                                         & {\bf{Definitions}}      \\
\hline
\hline
1.1 - Project Management       &      Management and systems engineering during the \\
& construction phase        \\
\hline
1.2 - Telescopes    &  Materials, equipment, labor, and travel associated with \\
& the design and construction of the telescopes   \\
\hline
1.3 - Telescope Receivers &  Materials, equipment, labor, and travel associated with \\
& the design, fabrication, assembly, and testing of the receivers\\
\hline
1.4 - Detectors and Readout &    Fabrication, assembly, and testing of the detectors and \\
& cold and warm readout electronics          \\
\hline
1.5 - Data Acquisition            &    Delivery of the control systems for the observatories \\
& and data acquisition    \\
\hline
1.6 - Data Management                         &   Maintenance of site computing,
networking and \\  
& data storage; staff for data acquisition,
pipeline \\
& development, and map making   \\
\hline
1.7 - Site Infrastructure &   Materials, equipment, labor, and travel needed to manage \\
& and oversee construction activities at the site             \\
\hline
1.8 - Integration and Testing    & On-site integration and commissioning of the telescopes\\
& and infrastructure         \\
\hline    

\hline
\end{tabular}
\label{tab:wbs}
\end{table}

\begin{table}[H]
\caption{{} Project Cost Estimates for CMB-HD Instrumentation and Construction — US Only {}}
\label{tab:my-table}
\centering
\begin{tabular}{|l|c|c|c|c|c|c|c|c|c|c|c|}
\hline
Project Item (WBS 1)                                        & Y1 & Y2 & Y3 & Y4 & Y5 & Y6 & Y7 & Y8 & Y9 & Y10 & Total (\$M)        \\
\hline
\hline
1.1 Project Management     & 10  & 10  & 10 & 10 & 10 & 10 & 10  & 10  & 10 & 10 & 100  \\
\hline
1.2 Telescopes       &   &   &  &  &  & 400 &   &   &  &  &      400   \\
\hline
1.3 Telescope Receivers &   &   &  & 20 & 20 & 20 & 20  & 20   &  &  &      100   \\
\hline
1.4 Detectors and Readout    &   &   &   & 64  & 64 & 64 & 64  & 64  &  &  &  320  \\
\hline
1.5 Data Acquisition   &   &   & 1 & 1 & 1 &1 & 1  &  1 & 1 & 1 &  8 \\
\hline
1.6 Data Management  & 2.4  & 2.4  & 2.4 & 2.4 & 2.4 &2.4 & 2.4  & 2.4  & 2.4 & 2.4 &   24  \\
\hline
1.7 Site Infrastructure &   &   &  & 10  & 10 & 6 & 6  & 6  & 6 & 6 &    50      \\
\hline
1.8 Integration and Testing  &   &   &  &  & 4 & 4 & 4  & 4  & 4 &  &   20  \\
\hline
Total  &   &   &  &  &  & &   &   &  &  &   1022      \\
\hline
\end{tabular}
\label{tab:cost}
\end{table}

\clearpage

\addcontentsline{toc}{section}{References}
\bibliographystyle{unsrturltrunc6}
\bibliography{cmb-in-hd.bib}

\clearpage
\section*{Affiliations}

\noindent $^{1}$ Stony Brook University \\
$^{2}$ Flatiron Institute \\
$^{3}$ École Normale Supérieure, Paris, France \\
$^{4}$ University of Bonn \\
$^{5}$ The University of Texas at Austin \\
$^{6}$ Arizona State University \\
$^{8}$ University of Louvain, Belgium \\
$^{9}$ University of Namur, Belgium \\
$^{10}$ Harvard University \\
$^{11}$ University of New Mexico \\
$^{12}$ Max Planck Institute for Astrophysics \\
$^{13}$ University of Pennsylvania \\
$^{14}$ NASA/Goddard Space Flight Center\\
$^{15}$ Lawrence Berkeley National Laboratory \\
$^{16}$ University of California, San Diego \\
$^{17}$ Pennsylvania State University \\
$^{18}$ University of Illinois at Urbana-Champaign \\
$^{19}$ Columbia University \\
$^{20}$ York University \\
$^{21}$ Perimeter Institute \\
$^{22}$ UK Astronomy Technology Centre \\
$^{23}$ University of Cambridge \\
$^{24}$ Southern Methodist University \\
$^{25}$ European Southern Observatory \\
$^{26}$ Institut D'Astrophysique De Paris \\
$^{27}$ Yale University \\
$^{28}$ Cambridge University \\
$^{29}$ University of California, Berkeley \\
$^{30}$ Cornell University \\
$^{31}$ University of Colorado \\
$^{32}$ University of Southern California \\
$^{33}$ Brookhaven National Laboratory \\
$^{34}$ University of Hamburg \\
$^{35}$ KICP University of Chicago

\end{document}